# AMD MI300X GPU Performance Analysis

Chandrish Ambati and Trung Diep

*Celestial AI*

## Abstract

The rapid growth of large language models (LLMs) has driven the need for high-performance, scalable GPU hardware capable of efficiently serving models with hundreds of billions of parameters. While NVIDIA GPUs have traditionally dominated LLM deployments due to their mature CUDA software stack and state-of the-art accelerators, AMD's latest MI300X GPUs offer a compelling alternative, featuring high HBM capacity, matrix cores, and their proprietary interconnect. In this paper, we present a comprehensive evaluation of AMD's MI300X GPUs across key performance domains critical to LLM inference: compute throughput, memory bandwidth, and interconnect communication.

We evaluate performance using vendor-provided microbenchmarks and supplement them with custom kernels where necessary to analyze compute, memory, and communication efficiency. We also provide a generational comparison of architectural features across both AMD and NVIDIA products, highlighting the gap between theoretical specifications and real-world performance. For full-system evaluation, we benchmark end-to-end inference throughput using the Llama 70B model under low-precision formats (FP8 and FP16). Our findings reveal that despite the MI300X offering higher theoretical compute capacity of 1.5x than that of the NVIDIA H100, the AMD GPUs achieve only 37–66% of H100/H200 performance in realized LLM inference performance.

## 1. Introduction

The rapid rise of large language models (LLMs) such as GPT, Llama, and Mistral has driven unprecedented demand for high-performance GPU hardware, not only to train these models, which can contain tens to hundreds of billions of parameters and take weeks to months to train, but also to serve them efficiently at scale for inference. The performance of LLM inference pipelines depends critically on multiple hardware dimensions: compute throughput (FLOPs), memory bandwidth and capacity, and interconnect performance (latency and bandwidth between GPUs). Equally important is a mature software ecosystem capable of fully leveraging these hardware capabilities.

While NVIDIA GPUs have long been the industry standard for production AI workloads supported by a robust and well-integrated CUDA software ecosystem there is growing momentum behind alternative hardware solutions, particularly from AMD.

In this white paper, we present a side-by-side evaluation of NVIDIA and AMD GPUs for LLM inference, focusing on three core dimensions of hardware performance:
- Compute: Measuring the number of floating-point operations per second (FLOPs) the hardware can do available for doing matrix multiplications
- Memory: Comparing the memory capacity and memory bandwidth available for storing and accessing weights and activations of the model
- Communication: Assessing the GPU interconnect technology over parameters of latency, bandwidth and topology

Our evaluation uses vendor-supported benchmarks and tooling, including frameworks such as vLLM and TensorRT-LLM, to assess real-world inference performance and system efficiency.

## 2. Performance Benchmarking of Compute

Tensor Cores (on NVIDIA GPUs) and Matrix Cores (on AMD GPUs) are specialized hardware accelerators designed to perform high-throughput matrix multiplications, which are the fundamental compute operations behind large language models (LLMs). A key metric for evaluating the raw computational capability of these units is the floating-point operations per second (FLOPs) they can sustain, especially for inference in low-precision formats like FP8, BF16, and FP16

*AMD Architecture:*
Earlier AMD GPUs architecture like GCN (Graphics Core Next) were built as general-purpose GPU (GPGPU) with same architecture for building solutions for PC Gamers and datacenter platforms. There are two architectures now: AMD RDNA optimized for gaming and maximizing the frames per second and AMD CDNA optimized for computing and push the limits of flops per second. CDNA-I compute units are built on the foundation of GCN architecture with enhancements to add Matrix Core Engines in each CU to boost computational throughput of fp64 and fp32. CDNA-2 architecture shows a giant leap from monolithic architectures to chiplet-based architecture enabling two GCDs (Graphics Compute Die) integrated into a single package in the OAM form factor in MI250 and MI250x products connected leveraging the AMD's unique on-die Infinity Fabric. CDNA-3 takes this architecture a step further with MI300 series products which can integrate up to 8 vertically stacked accelerator compute dies (XCD). MI300x comes with 8 XCD dies almost doubling the computational capability from previous generation and MI300A APU is three Zen 4 x86 CPU dies integrated with 6 XCDs in a single package.

|  | B200 | H100 | MI300X | MI300A | MI250X | MI250 | MI210 | MI100 |
|---|---|---|---|---|---|---|---|---|
| CUs/ SMs | 160 | 132 | 304 | 228 | 220 | 208 | 104 | 120 |
| tf32 | 1100 | 495 | 654 | 490 | - | - | - | - |
| fp16/bf16 | 2250 | 989 | 1307 | 981 | 383 | 362 | 181 | 185 |
| fp8 | 4500 | 1979 | 2615 | 1961 | - | - | - | - |
| int8 | 4500 | 1979 | 2615 | 1961 | 383 | 362 | 181 | 185 |
| fp32 | 75 | 67 | 163 | 123 | 96 | 91 | 45 | 46 |
| fp64 vector | 37 | 34 | 82 | 61 | 48 | 45 | 23 | 12 |
| fp64 matrix | 37 | 67 | 163 | 123 | 96 | 91 | 45 | - |

*Table 1: Peak Theoretical TFLOPs (dense)*

## 2.1. Benchmarking Methodology

To evaluate the matrix multiplication performance, we measured each platform using its most optimized and actively maintained linear algebra libraries. For AMD GPUs, we utilized the ROCm software stack, which includes several BLAS libraries such as rocBLAS, hipBLAS, and hipBLASLt. Among these, we selected hipblaslt-bench as the primary benchmarking tool for the MI300X platform. This choice was motivated by its active development, as well as its support for newer datatypes such as FP8, which are not currently available in the other AMD libraries.

For NVIDIA GPUs, we used cuBLASLt, part of NVIDIA's CUDA toolkit, which provides highly optimized kernels for general matrix multiplications (GEMMs). To measure performance on H100 (Hopper) and B200 (Blackwell) GPUs, we leveraged the cublaslt_gemm microbenchmark within Microsoft's Superbench, a widely used validation and profiling suite for AI infrastructure.

In both cases, we benchmarked square matrix multiplications by fixing the matrix dimensions such that M=N=K. The tested sizes ranged from 64 to 65,536, covering both small and extremely large GEMMs. We primarily used the NT memory layout (i.e., input A in row-major, input B in column-major) for consistency across platforms.

## 2.2. Results

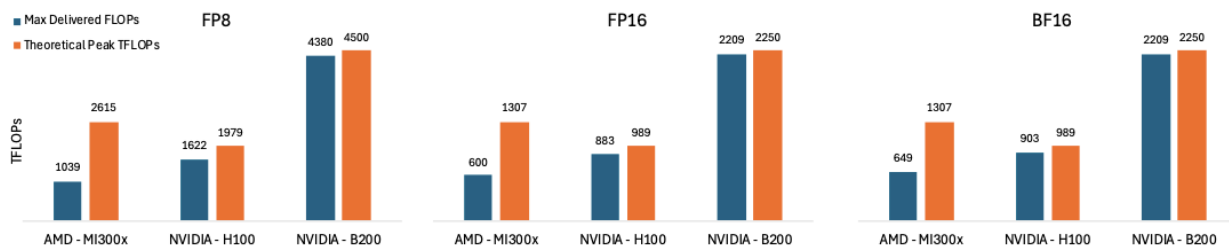

*Figure 1: Theoretical vs Max Delivered FLOPs*

From Figure 1, we observe that the AMD MI300X can achieve, on average, only 45% of its theoretical peak FLOPs across FP8, BF16, and FP16 precision formats which is cross referenced with AMD published blog[]. In contrast, NVIDIA's H100 and B200 GPUs can sustain up to 93% of their peak compute throughput, indicating significantly higher utilization of available hardware performance.

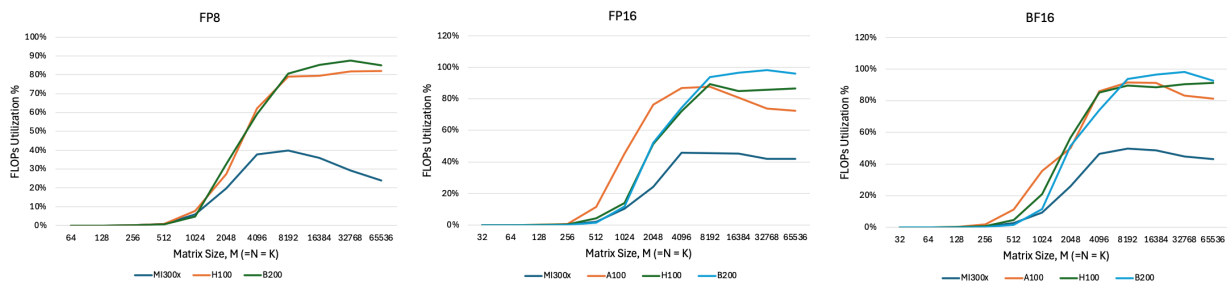

*Figure 2: FLOPs Utilization % vs Matrix Sizes, Datatype: fp8, bf16, fp16*

Both the NVIDIA GPUs, H100 and B200 exhibit strong scaling behavior achieving over 90% of the theoretical FLOPs at moderate matrix sizes (~4096 and above). MI300x underperforms peaking up to 50% utilization for size of 4096 and above. We also notice that there is slight dip in performance with larger matrix sizes. This phenomenon can be attributed to the memory bound kernels for these large matrices. Additionally, large problem sizes often exceed the effective capacity of the shared memory and caches, hence resulting in more and frequent memory accesses to slower global memory.

## 2.3. Software Efficiency

This large disparity between theoretical and observed FLOPs on the MI300X can be primarily attributed to two factors: (1) software stack efficiency, including compiler and kernel maturity, and (2) dynamic frequency scaling and power management behavior, which may limit clock speeds under load. The theoretical peak FLOPs for each GPU are estimated using the following equation:

Peak FLOPs = Max Boost Clock Frequency x Number of Cores x $\frac{\text{OPS}}{\text{Core. Cycle}}$

To better isolate the effects of software overhead and frequency scaling, we conducted a microbenchmark experiment using matrix sizes designed to align closely with the number of compute units. We simultaneously recorded the sustained clock frequencies and observed throughput, allowing us to more precisely quantify the contributions of each limiting factor. The results of this validation experiment are shown below.

| Datatype | (M, N, K) | Boost Clock Freq (MHz) | Measured Clock Freq | Ops/Core/Cycle | Measured TFLOPs | Calculated Peak TFLOPs | Software Efficiency |
|---|---|---|---|---|---|---|---|
| fp8 | (4096, 3648, 32896) | 2100 | 1217 | 4096 | 1228 | 1515 | 81% |
| fp16 | (4096, 4864, 32896) | 2100 | 1083 | 2048 | 573 | 674 | 85% |
| bf16 | (4096, 4864, 32896) | 2100 | 1187 | 2048 | 588 | 739 | 80% |

*Table 2: Measured Clock Frequency and Software Efficiency*

## 3. Performance Benchmarking of Memory

Modern deep learning workloads like large-scale transformer architectures are heavily memory bandwidth bound. As model sizes and sequence lengths continue to grow, feeding data to compute units efficiently becomes a primary performance bottleneck. High Bandwidth Memory (HBM) is a state-of-the-art DRAM technology engineered to deliver extremely high memory bandwidth, reaching into the terabytes-per-second range. HBM achieves this by vertically stacking multiple DRAM dies, interconnected using through-silicon vias (TSVs) and mounted on an interposer alongside the compute die (typically a GPU). This integration enables wide and energy-efficient memory interfaces, significantly reducing latency and power per bit transferred compared to traditional GDDR or DDR memory systems.

The number of HBM stacks that can be mounted is physically constrained by the substrate area surrounding the compute die. For instance, AMD's MI300X supports eight HBM stacks, while NVIDIA's H100 includes six stacks, one of which may be reserved for redundancy or yield purposes. Each stack's total memory capacity depends on two key factors: a) the stack height i.e., the number of memory layers (dies) in the vertical stack and b) the per-layer DRAM density. Meanwhile, memory bandwidth is primarily determined by a) the number of data pins in the interface and b) the per-pin signaling speed. The evolution of HBM across generations has been driven by both capacity and bandwidth scaling. accomplished through innovations such as transitioning signaling technologies from NRZ (Non-Return to Zero) to PAM4 (4-level Pulse Amplitude Modulation), effectively doubling per-pin data rates or adoption of Hybrid Copper Bonding (HCB) to increase stack height and reduce TSV resistance.

These advances have allowed modern accelerators like the MI300X and H100 to deliver unprecedented memory performance. Table below summarizes HBM versions, stack counts, and memory specifications across recent AMD and NVIDIA architectures.

|  | MI 355x | MI 325X | MI 300X | MI300 A | MI250X | MI250 | MI210 | MI100 |
|---|---|---|---|---|---|---|---|---|
| Arch | CDNA 4 | CDNA3 | CDNA3 | CDNA3 | CDNA2 | CDNA2 | CDNA2 | CDNA1 |
| Memory | 288 GB | 256 GB | 192 GB | 128 GB | 128 GB | 128 GB | 64 GB | 32 GB |
| HBM ver. | HBM3e | HBM3e | HBM3 | HBM3 | HBM2e | HBM2e | HBM2e | HBM2 |
| Mem BW | 8 TB/s | 6 TB/s | 5.3 TB/s | 5.1 TB/s | 3.2 TB/s | 3.2 TB/s | 1.6 TB/s | 1.2 TB/s |
| # stacks | 8 | 8 | 8 | 8 | 8 | 8 | 4 | 4 |
| Stack | 36 GB | 32 GB | 24 GB | 16 GB | 16GB | 16 GB | 16 GB | 8 GB |

*Table 3: AMD GPUs HBM Memory*

|           | GB200     | B200      | H200    | H100 NVL | H100 SXM | A100    |
|-----------|-----------|-----------|---------|----------|----------|---------|
| Arch      | Blackwell | Blackwell | Hopper  | Hopper   | Hopper   | Ampere  |
| Memory    | 186 GB    | 180 GB    | 141 GB  | 94 GB    | 80 GB    | 80 GB   |
| HBM ver.  | HBM3e     | HBM3e     | HBM3e   | HBM3     | HBM3     | HBM2e   |
| Mem BW    | 8 TB/s    | 7.7 TB/s  | 4.8 TB/s| 3.9 TB/s | 3.35 TB/s| 1.9 TB/s|
| # stacks  | 8         | 8         | 6       | 6        | 5        | 5       |
| Mem/stack | 24 GB     | 24 GB     | 24 GB   | 16 GB    | 16 GB    | 16 GB   |

*Table 4: NVIDIA GPUs HBM Memory*

## 3.1. Benchmarking Methodology

To evaluate peak memory bandwidth performance, we used two distinct approaches tailored to each GPU vendor. For AMD GPUs, we utilized BabelStream, a synthetic benchmark based on the STREAM benchmark for CPUs. BabelStream is specifically designed to measure memory transfer rates between high-bandwidth memory (HBM) and compute units, and supports backends such as HIP, OpenMP, and Kokkos. In our study, we compiled for the HIP backend to benchmark AMD's MI300X GPUs.

BabelStream operates on three large arrays a, b, and c, with an optional scalar alpha. Each kernel processes N elements, and bandwidth is calculated based on the total number of bytes read and written ($\beta$ = bytes per data element):

| Operation | Kernel | Bytes Read/Written |
|-----------|--------|--------------------|
| Copy      | c[i] = a[i] | $2N\beta = N\beta$ reads + $N\beta$ writes |
| Multiply  | b[i] = alpha * c[i] | $2N\beta = N\beta$ reads + $N\beta$ writes |
| Add       | c[i] = a[i] + b[i] | $3N\beta = 2N\beta$ reads + $N\beta$ writes |
| Triad     | a[i] = b[i] + alpha * c[i] | $3N\beta = 2N\beta$ reads + $N\beta$ writes |
| Dot       | sum += a[i] * b[i] | $2N\beta = 2N\beta$ reads (no writes) |

For NVIDIA GPUs, we developed a custom CUDA kernel that mirrors BabelStream's memory access patterns, focusing on the Copy kernel operation. This allowed us to ensure consistency in the benchmarking methodology while adapting to the CUDA programming environment and leveraging NVIDIA-specific optimizations.

## 3.2. Kernel Implementation

For the MI300X, we compiled and executed the HIP-based implementation of BabelStream. In this configuration, each thread is responsible for processing a single data element. The kernel was launched with a fixed thread block size of 1024, and the total number of thread blocks was set to N/1024, where N is the total number of elements in the array. This configuration is designed to ensure full utilization of the GPU's compute resources while maximizing sustained memory throughput for large-scale array operations.

To enable a fair comparison with the AMD setup, we implemented a custom Copy kernel in CUDA for benchmarking memory bandwidth on NVIDIA GPUs. In this kernel, each thread reads a single element from array A and writes it to array B. We evaluated several thread block sizes—256, 320, 512, and 1024—to determine optimal performance. We selected 320 threads per block, aligning with the internal configuration commonly used by NCCL for its collective communication kernels. The grid size was determined dynamically to ensure full coverage of the dataset, with the number of thread blocks calculated based on the total number of elements being processed.

## 3.3. Results

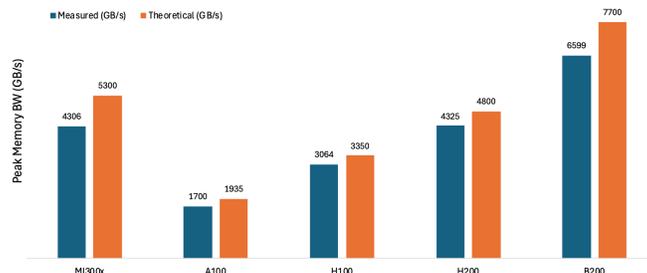

*Figure 3: Theoretical vs Measured Peak Memory BW*

The figure above presents a comparison between theoretical and measured peak memory bandwidth for five GPUs commonly used in LLM inference workloads: AMD MI300X and NVIDIA A100, H100, H200, and B200. The MI300X exhibits strong memory performance, achieving approximately 81% of its theoretical peak of 5.3 TB/s, indicating a well-utilized memory subsystem. NVIDIA's earlier architectures A100, H100, and H200 demonstrate even higher utilization, with measured bandwidths reaching up to 90% of their respective theoretical maximum. This reflects the maturity of NVIDIA's memory controller design and kernel stack optimizations. Notably, the B200 reaches about 86% of its 7.7 TB/s theoretical limit, suggesting that kernel-level tuning is still in progress. This was confirmed by re-running benchmarks one month later, which showed a 10% improvement in measured bandwidth, pointing to ongoing software level optimization efforts by NVIDIA for Blackwell.

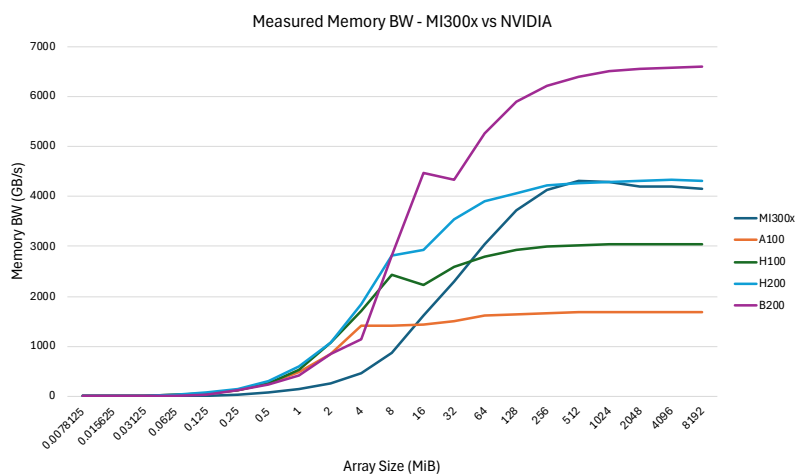

*Figure 4: Memory Bandwidth of GPUs across Array Sizes*

Figure 4 illustrates the scaling of measured memory bandwidth across a range of array sizes for the AMD MI300X and NVIDIA A100, H100, H200, and B200 GPUs. The results highlight that NVIDIA's B200 leads all architectures, achieving a peak bandwidth of approximately 6.6 TB/s with sustained scaling beyond 512 MiB array sizes. The MI300X shows competitive performance, saturating at around 4.3 TB/s, though its bandwidth plateaus earlier—around 64–128 MiB. The H200 slightly outperforms the H100, suggesting incremental architectural improvements in the HBM memory technology. In contrast, the A100 saturates early at around 1.7 TB/s, reflecting the limitations of its older HBM2 memory stack. Overall, these results emphasize the importance of memory bandwidth scaling in large-model inference workloads and showcase the efficiency advantages of newer GPU architectures when handling large activation and weight tensors typical of LLM inference.

## 4. Performance Benchmarking of Communication

Up to this point, we have examined the compute and memory characteristics of individual GPUs in isolation. However, as the size of modern large language models continues to grow, they increasingly exceed the capacity of a single GPU, primarily due to limitations in available HBM memory. As discussed in Section 3, the ability to scale HBM capacity and bandwidth is constrained by both physical memory density limitations and the technological complexity of advancing high-bandwidth memory interfaces. Consequently, to run such models efficiently, workloads must be distributed across multiple GPUs, which places significant importance on the efficiency of inter-GPU communication. How these GPUs are interconnected both in terms of bandwidth and latency becomes a critical factor influencing overall system performance, especially for tensor-parallel or pipeline-parallel inference and training workloads.

NVIDIA's NVLink is a high-speed, low-latency interconnect designed to enable scale-up GPU architectures, facilitating rapid data transfer both between GPUs and between GPUs and CPUs. In newer architectures such as Grace Blackwell (GB200), NVLink serves as a foundational element of NVIDIA's unified memory and compute fabric. Each generation of NVLink has delivered substantial improvements in bandwidth and interconnect efficiency. The latest fifth-generation NVLink offers up to 1800 GB/s of bidirectional bandwidth per GPU for collective communication operations such as all-reduce. In this study, we evaluate NVLink 4.0, which is used in systems based on the Hopper (H100) architecture.

On the AMD side, Infinity Fabric (IF) serves as the company's proprietary interconnect for both intra-GPU communication linking multiple compute dies (GCDs) within a single MI300X device and inter-GPU communication across devices in a node. The most recent generation of Infinity Fabric supports bidirectional bandwidths of up to 128 GB/s per link, enabling coherent memory access and synchronized compute across the distributed chiplet-based architecture.

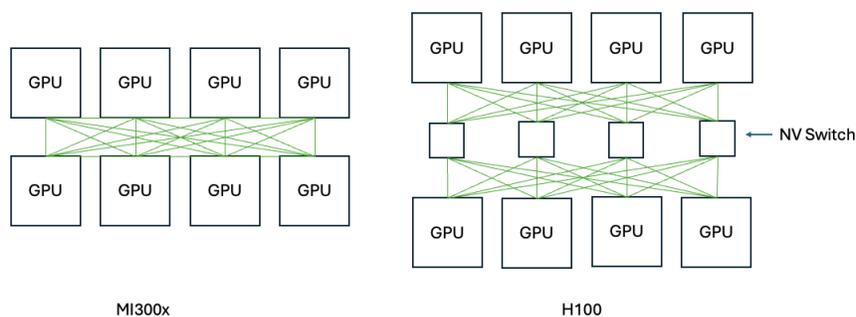

*Figure 5: Logical Representation of Scale Up Topologies of MI300x and H100*

### 4.1. Benchmarking Methodology

NVIDIA Collective Communications Library (NCCL) is a high-performance library developed by NVIDIA to enable optimized collective communication (e.g., all-reduce, broadcast, all-gather) between GPUs. It is topology-aware, meaning it adapts its algorithms based on the system's NVLink, NVSwitch, or PCIe layout to maximize bandwidth and minimize latency. RCCL (ROCm Collective Communication Library) serves as AMD's counterpart for multi-GPU communication. It is built as a fork of NCCL, maintaining similar APIs and design principles, but tailored for AMD's hardware, including support for Infinity Fabric-based interconnects and the ROCm software stack.

To evaluate inter-GPU communication performance, we utilize NCCL-tests, a widely adopted benchmarking suite designed to assess both the performance and correctness of collective communication operations implemented via the NCCL. Our experiments include a range of collectives—all-reduce, all-gather, all-to-all, broadcast, and reduce-scatter—executed across multiple GPUs. Message sizes are varied from 8 bytes up to 4 GB, using single-precision floating-point data types, to simulate realistic deep learning communication patterns. For benchmarking on AMD GPUs, we employ RCCL-tests, which is based on the same codebase and performance evaluation methodology as NCCL-tests and serves as the corresponding tool for the RCCL. This ensures consistency in how communication metrics are collected and allows for a direct performance comparison between NVIDIA and AMD GPU interconnects.

## 4.2. Results

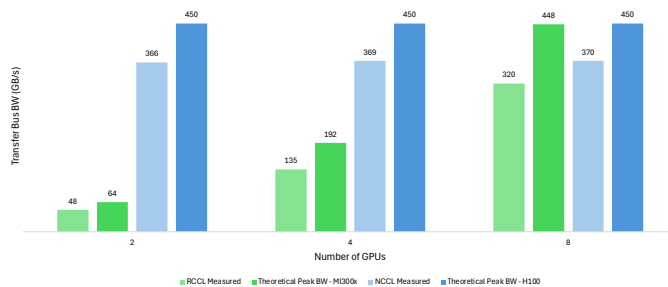

*Figure 6: Inter GPU Communication Bandwidth*

Figure 5 illustrates the measured peak inter-GPU bandwidth during collective communication operations—specifically all-reduce—across configurations with 2, 4, and 8 GPUs. The evaluation compares NVIDIA H100 GPUs, connected via NVLink 4.0 (NV18) with a bidirectional bandwidth of 900 GB/s, against AMD MI300X GPUs, interconnected through Infinity Fabric links offering 128 GB/s bidirectional bandwidth per link.

In the 8-GPU NVIDIA DGX H100 system, GPUs are connected using a NVSwitch topology as shown in Fig 5, enabling uniform and consistent bandwidth regardless of the number of GPUs involved. In contrast, the 8-GPU AMD MI300X platform employs a mesh-based Infinity Fabric topology, where aggregate bandwidth scales with the number of active GPU pairs and communication links.

Measured results show that NVIDIA GPUs achieve approximately 85% of their theoretical peak bandwidth, reflecting a mature and highly optimized collective communication stack in NCCL. For AMD, bandwidth scales from 64 GB/s (2 GPUs) to 192 GB/s (4 GPUs) and 448 GB/s (8 GPUs), reaching about 70% of theoretical peak bandwidth. There remains headroom for improvement particularly in optimizing RCCL to better exploit AMD's mesh interconnect layout.

From an LLM inference perspective, these interconnect characteristics have a direct impact on tensor parallelism (TP) communication, especially when model partitions span multiple GPUs. Enhancing AMD's collective algorithms to match the topology-aware strategies used in NCCL would significantly reduce communication overhead and improve scalability for large-scale inference deployments.

## 5. LLM Performance

While microbenchmarks are valuable for isolating and comparing the core hardware components such as compute, memory, and communication bandwidth—that impact large language model (LLM) inference, they do not capture the full complexity of end-to-end execution. To address this, we conducted comprehensive end-to-end inference experiments using Llama 3.1 70B as a representative model. These tests were performed on NVIDIA H100 and H200 GPUs as well as on AMD MI300X, utilizing rapidly evolving software stacks optimized for each platform.

To evaluate real-world performance, we tested across a wide range of input and output sequence lengths, capturing both short-form and long-form generation use cases. For NVIDIA GPUs, we used the TensorRT-LLM benchmark suite, while for AMD GPUs, we leveraged the vLLM framework with support for both FP8 and FP16 inference. These experiments reflect the combined influence of hardware capability and software stack maturity on actual inference throughput.

## 5.1. FP8 Results

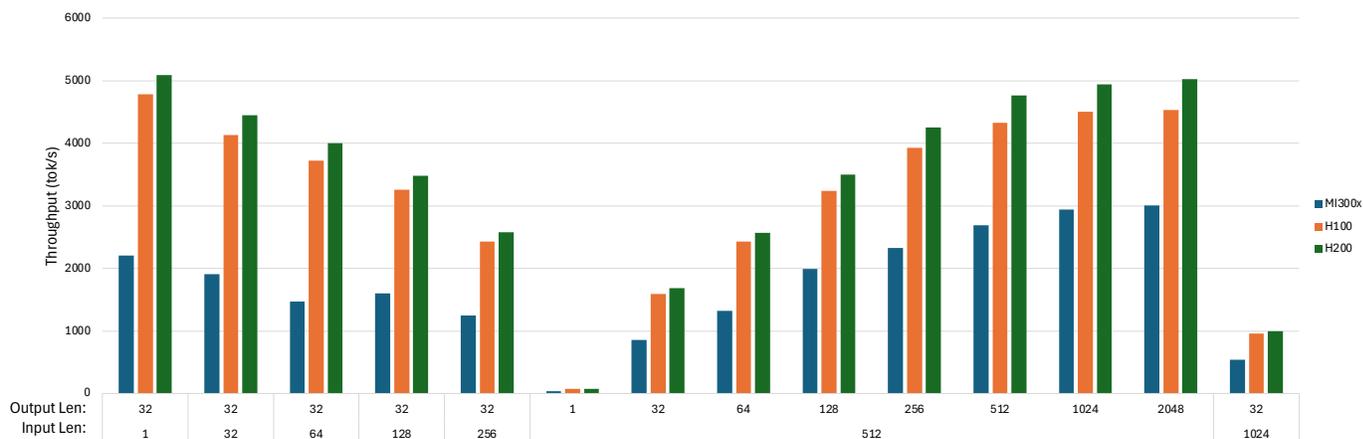

*Figure 7: Llama 3.1 70B - fp8 comparison*

Figure 6 presents the inference throughput (tokens/second) for the Llama 3.1 70B model in FP8 precision across three GPUs: NVIDIA H100, NVIDIA H200, and AMD MI300X. LLM inference is composed of two primary phases:
- Prefill phase: A compute-bound stage where attention is calculated over the entire input sequence.
- Decode phase: A memory-bound stage where key-value (KV) cache reuse dominates, especially at longer output lengths.

The throughput is computed using the following equation:
$$\text{Throughput (tok/s)} = \frac{\text{Output Tokens}}{\text{Prefill Time} + \text{Decode Time}}$$

*Prefill-Dominated Regime (Input Len ↑, Output Len = 32)*:
As the input length increases from 32 to 256 with a fixed short output length, throughput decreases across all GPUs. This trend is due to the rising prefill cost, which increases the denominator of the throughput equation. In this regime, MI300X consistently delivers only around 50% or less of the throughput achieved by H100 and H200, which aligns with the earlier discussion on MI300X's lower performance for matrix multiplications in the compute section of this paper.

*Decode-Dominated Regime (Input Len = 512, Output Len ↑):*
When input length is fixed at 512 and output length increases from 1 to 2048, throughput steadily rises for all GPUs. This is expected, as the numerator (number of output tokens) increases faster than the growth in decode time. Here, memory bandwidth becomes the dominant factor, and MI300X closes the performance gap, improving from 49% of H100's performance (at output length 1) to 66% (at output length 2048). This is consistent with the competitive memory bandwidth performance of MI300X, as discussed in the memory section of this paper.

## 5.2. FP16 Results

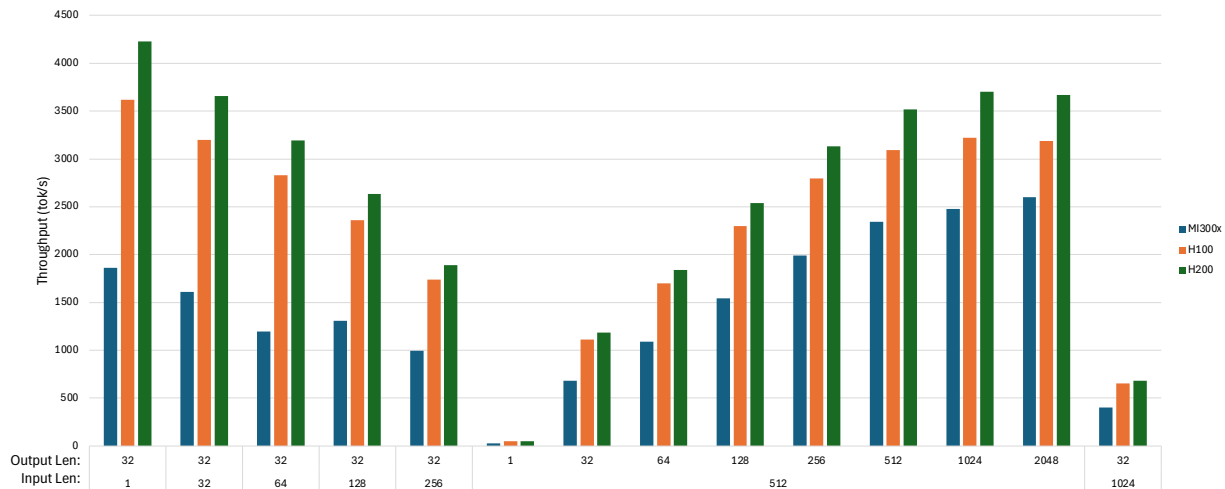

Figure 8: Llama 3.1 70B - fp16 comparison

A similar trend to the FP8 results is observed in the FP16 inference throughput graph, Fig 8. Notably, the MI300X demonstrates improved relative performance in the decode-dominated regime, where its throughput increases from approximately 56% to 80% of the NVIDIA H100's performance. This improvement surpasses the corresponding FP8 results and can be explained through the memory bandwidth characteristics shown in Figure 6. Specifically, for smaller working set sizes (1–64 MB for H100 and up to 128 MB for H200), the MI300X exhibits lower effective bandwidth compared to NVIDIA GPUs. These memory sizes are representative of typical FP8 inference workloads, where the model weights and KV cache footprints fall within this lower-bandwidth region. Since the decode phase of inference is memory-bound, the lower bandwidth directly impacts MI300X's performance in FP8 scenarios.

In contrast, FP16 inference doubles the memory footprint of model weights and KV caches, shifting the workload into the higher range of the bandwidth curve, where MI300X achieves comparable or superior memory throughput relative to H100 and H200. This shift results in a narrowed performance gap during decode, highlighting the impact of memory system behavior on inference performance across precision formats.

## 6. Summary

While AMD has made substantial progress in advancing the hardware specifications of its GPU offerings, realizing the full potential of these capabilities is contingent upon a mature and robust software ecosystem. Despite hardware innovations such as the introduction of the Infinity Fabric 128, AMD's software stack currently lags NVIDIA's, which continues to evolve rapidly with frameworks like NVIDIA Dynamo announced at recent GTC conferences.

Although AMD GPUs have demonstrated value in research environments such as their deployment in Frontier, the exascale system at Oak Ridge National Laboratory broader adoption in production LLM workloads will require consistent performance and support across the entire software stack, not just isolated benchmarks.

Key areas for improvement include accelerating the development of collective communication libraries and interconnect technologies. While NVIDIA's NCCL is integrating support for NVSHMEM, AMD's software ecosystem would benefit from incorporating similar high-performance communication primitives to improve scalability and efficiency in multi-GPU inference. Bridging these software gaps will be essential for AMD to become a competitive platform for large-scale, production-grade AI workloads.